\documentclass[11pt]{article}
\usepackage{amssymb}
\usepackage{graphicx}
\usepackage{amsmath}
\usepackage[all]{xy}
\newcommand{\be}{\begin{equation}}
\newcommand{\ee}{\end{equation}}
\newcommand{\bqa}{\begin{eqnarray}}
\newcommand{\eea}{\end{eqnarray}}
\newcommand{\bqas}{\begin{eqnarray*}}
\newcommand{\eeas}{\end{eqnarray*}}

\newcommand{\non}{\nonumber}

\def\brc{\langle}
\def\ckt{\rangle}

\def\D{\mathcal{D}}
\def\Tr{\hbox{\rm Tr}}

\def\Tr{ \hbox{\rm Tr}}

\def\im{\hbox{\rm Im}}

\usepackage{bbm}

\begin{document}

\thispagestyle{empty}

\begin{flushright}
 IFUP-TH/2008-24, \\
{\tt hep-th/yymmnnn} \\
September, 2008 \\
\end{flushright}
\vspace{3mm}

\begin{center}
{\Large \bf
Vortices which do not Abelianize dynamically: \\
Semi-classical origin of non-Abelian monopoles }
\footnote{ Contribution to ``Continuous Advances in QCD 2008'',  Minneapolis, May 2008. To appear in the Conference Proceedings.  } 
\\[12mm]
\vspace{5mm}

\normalsize
  { \bf
Kenichi Konishi}
\footnote{e-mail: konishi(at)df.unipi.it}

\vspace{3mm} 

{\it Department of Physics, ``E. Fermi'',  University of Pisa \\
and 
\\
 INFN, Sezione di Pisa, \\
Largo Pontecorvo, 3, Ed. C, 56127 Pisa, Italy
} \\
%
%
%
\vspace{5mm}

{\bf Abstract}  \\[5mm]

{\parbox{13cm}{\hspace{5mm}
After briefly reviewing the problems associated with non-Abelian monopoles, we turn our attention to the development in our understanding of non-Abelian {\it vortices} in the last several years.  In the $U(N)$  model  with $N_{f}=N$ flavors  in which they were first found,  the fluctuations of the orientational modes along the vortex length and in time become strongly coupled at long distances.  They effectively reduce to Abelian ANO vortices.  We discuss then a very recent work on non-Abelian  vortices  with $CP^{n-1}\times CP^{r-1}$ orientational moduli,  which, unlike the ones so far extensively studied,  do not dynamically Abelianize completely.  The surviving vortex orientational moduli, fluctuating along the vortex length and in time,  gets absorbed by the  monopoles at the ends, turning into the dual gauge degrees of freedom for the latter.

}}
\end{center}

\newpage

\section{Non-Abelian monopoles}

Non-Abelian monopoles have been introduced as a natural generalization of the 't Hooft-Polyakov monopoles \cite{TH}; they  arise in systems with partial gauge symmetry breaking, 
\be
 G   \,\,\,{\stackrel {v_{1}} {\longrightarrow}} \,\,\, H  \,\,
\ee 
where $H$ is a non-Abelian gauge group. 
The regular monopoles arising in such a system are characterized by the charges $\beta$  such that 
\begin{equation}   F_{ij} =  \epsilon_{ijk} B_k = 
\epsilon_{ijk}  \frac{ r_k 
}{      r^3}  ({ \beta} \cdot  {\bf H}),           \end{equation}
in an appropriate gauge, where ${\bf H}$ are the diagonal generators of $H$ in the Cartan subalgebra. 
A straightforward generalization of the Dirac's quantization condition leads to 
\begin{equation}  2 \, {\beta \cdot \alpha} \in  { \bf Z}   \label{naqcond}
\end{equation}
where $\alpha$ are the root vectors of $H$.
The constant vectors $\beta$  (with the number of components equal to the rank of the group $H$)  label possible monopoles. 
It is easy to see that the solution of Eq. (\ref{naqcond})  is  that  $\beta$  is any  of the {\it weight vectors} of a group whose  
nonzero roots are given by 
\be   \alpha^{*} = {\alpha}/{\alpha \cdot \alpha}. 
\label{dualg}\ee
 The group generated by Eq. (\ref{dualg}) is known as the (GNOW)  dual of  $H$,  let us call ${\tilde H}$.  One is thus led to  a set of semi-classical {\it degenerate}  monopoles, with multiplicity   equal to that of a representation of ${\tilde H}$;  this has led to the so-called  GNOW   conjecture, {\i.e.}, that they form a multiplet of the group ${\tilde H}$,   dual of  $H$ \cite{GNO}.   
 For simply-laced groups (with the same length of all nonzero roots) such as $SU(N)$, $SO(2N)$, the dual of $H$ is basically the same group, 
except that the allowed representations tell us that 
\be     U(N) \leftrightarrow  U(N);  \qquad   SO(2N)\leftrightarrow  SO(2N) ,  \ee
while 
\be  SU(N) \leftrightarrow  {SU(N)}/{{\mathbf Z}_{N}}; \qquad   SO(2N+1)   \leftrightarrow  USp(2N).   \label{below}
\ee

  There are however well-known difficulties with such an interpretation.  The first concerns the topological obstruction discussed
in \cite{CDyons,DFHK}:  in the presence of the classical monopole background, it is not possible  to define a globally  well-defined set of generators isomorphic to $H$.  As a consequence, no  ``colored dyons''  exist.  In a simplest case with 
the breaking 
\be  SU(3)  \,\,\,{\stackrel {\brc \phi_{1} \ckt    \ne 0} {\longrightarrow}} \,\,\, SU(2) \times U(1),
\label{simplebr}\ee
   this means  that 
    \be { no\,\, monopoles\,\, with \,\, charges }  \quad   ({\underline 2},  1^{*})  \quad  { exist},    \label{cannot} \ee
   where the asterisk indicates a dual, magnetic charge.  

The second can be regarded as an infinitesimal version of the same difficulty:   certain bosonic zero modes around the monopole solution, corresponding to $H$ gauge transformations,  are non-normalizable (behaving as $r^{-1/2}$ asymptotically).  Thus the standard procedure of quantization leading to  $H$ multiplets of monopoles  does not work \cite{DFHK}.   

Both of these difficulties concern  the transformation properties of the  monopoles  under the subgroup  $H$,   while the  relevant question should be how they transform under the dual group, ${\tilde H}$.  As field transformation groups, $H$ and ${\tilde H}$  are relatively nonlocal, the latter  should look like a nonlocal transformation group  in the original, electric description.

\section{Light non-Abelian monopoles}

In spite of these apparent difficulties, light non-Abelian monopoles do appear regularly in the low-energy effective action of a wide class of ${\cal N}=2$ gauge theories with matter hypermultiplets \cite{APS,HO,CKM}.  It is however important to bare in mind that  non-Abelian dual gauge groups (and associated monopoles) occur only in models with massless  flavors in the underlying theory.    The renormalization-group effects explain this fact.   For instance, in the softly broken, ${\cal N} =2$  supersymmetric $SU(N)$ ($N \ge 3$)  gauge theory with $N_{f}$ quarks, 
the low-energy, magnetic group
$SU(r) \times U(1)^{N-r}$  appears  only for 
\[    r \le   {N_{f}}/{2}\;.
\]
The reason is that the monopoles, via the Jackiw-Rebbi mechanism,  form a degenerate flavor multiplet quantum mechanically.  (Indeed the light monopoles appear in the above theory as a fundamental multiplet of the flavor $SU(N_{f})$ symmetry group.)   Their quantum effects attenuate the dual gauge interactions so that the $SU(r)$ dual gauge group is now infrared-free  (the sign flip with respect to the underlying theory \cite{Konishi04}).   This is how these objects can appear in the infrared as a recognizable degrees of freedom.  
When this is not possible (e.g.,  pure ${\cal N}=2$  gauge theories)  the would-be non-Abelian monopoles interact too strongly and form (baryon-like) composites, which are the 
Abelian monopoles \cite{CKM}.

The vacua $r= N_{f}/2$  constitute interesting, limiting class of theories: they are infrared fixed-point theories (SCFT) \cite{AD}. 

  Now there must be ways to {\it understand} these massless non-Abelian monopoles,  in spite of the above-mentioned difficulties, in terms of more familiar, semiclassical language.  Below we shall show that this is indeed possible.    We shall study the monopoles in terms of {\it  vortices}, by putting the low-energy $H$  gauge system in Higgs phase.  A systematic study of non-Abelian {\it vortices} started only recently,  but  are much better understood than the non-Abelian {\it monopoles}.  The monopoles and vortices are closely  related to each other,  through the homotopy map and by symmetry \cite{Konishi04,ABEK,Duality}.    The moduli and non-Abelian transformation properties among the monopoles follow from those of the low-energy vortices which confine them.

\section{Non-Abelian vortices \label{aba:sec1}}

The non-Abelian vortices have been found several years ago \cite{HT,ABEKY},  in the context of $U(N)$
gauge theory with $N_{f}$ flavors, $N_{f}\ge N$.\footnote{Recently the construction has been generalized to any gauge group of the form, $G= G^{\prime}\times U(1)$ \cite{AGG}.   }     Even though such a model can be considered in its own right,  we shall follow here the original approach \cite{ABEKY}  where the  $U(N)$  model arises as the low-energy approximation after a partial symmetry breaking,  
\be    SU(N+1)  \,\,\,{\stackrel  {v_{1}    \ne 0} {\longrightarrow}} \,\,\, \frac{ SU(N) \times U(1)}{{\mathbf Z}_{N}}\sim U(N)\;. 
\label{above}  \ee
This approach allows us to connect the vortex properties to those of the monopoles appearing at the ends, as neither the vortices nor monopoles (arising from the symmetry breaking Eq.~(\ref{above}))  are truly stable \cite{ABEK,Duality,Konishi04}.   The point is that at a much smaller mass scale, $v_{2}\ll v_{1}$, the squark fields condense, and break the low-energy $U(N)$ gauge symmetry completely.

At scales much lower than  $v_{1} $ but still neglecting the smaller squark  VEV
$v_{2}$   the theory reduces to an $SU(N)\times U(1)$  gauge theory \cite{ABEKY}  with $N_{f}$ light quarks $q_{i}, {\tilde q}^{i}$ (the first $N$ components of the original quark multiplets $Q_{i}, {\tilde Q}^{i}$).    In the most frequently studied case,   $N_{f}=N$, the light squark fields can be expressed as $N\times N$  color-flavor  mixed matrix.   
 The adjoint scalars are fixed to its VEV of the form, 
 \[    \brc \Phi \ckt = {\rm diag}\, (m,m, \ldots, m, - N\, m), 
 \]
  with small fluctuations around it,
\be  \Phi =  \brc\Phi  \ckt  (1 +    \brc\Phi  \ckt^{-1} \, {\tilde  \Phi} )  , \qquad   |{\tilde  \Phi}| \ll m.
\label{small}\ee
In the consideration of the vortices of the low-energy theory,  $\Phi$  will be in fact replaced by the constant VEV.  The presence of the small terms Eq. (\ref{small}), however, makes the low-energy vortices not strictly BPS  (and this is  important in the consideration of their stability).

The quark fields are replaced,  consistently with the vanishing of the D-term potential,  as
\be    {\tilde q} \equiv   q^{\dagger}, \qquad   q \to  \frac{1}{\sqrt{2}} \, q,
\ee
where the second replacement brings back the kinetic term to the standard form.

We further replace  the singlet coupling constant and the $U(1)$  gauge field 
appropriately: 
the net effect is
\be  {\cal L} =  \frac{ 1}{ 4 g_N^2}  (F_{\mu \nu}^a)^2  + \frac{ 1}{ 4 e^2}  ({\tilde F}_{\mu \nu})^2  +
 \left|{\cal D}_{\mu}
q \right|^2
-    \frac{e^2}{2} \, |
 \, q^{\dagger} \,  q   -     c \, {\mathbf 1}  \, |^2 - \frac{1}{2} \, g_N^2 \,| \,
 \,  q^{\dagger} \, t^a q \,  |^2,
\ee
where
$  c=   N(N+1)  \sqrt {    2\, {  \mu \, m}    }$. 
Neglecting the small  terms left implicit, this  is basically the $U(N)$  model,  studied extensively \cite{HT,ABEK,Etou,Eto:2006pg,Tong,SY,GSY,Duality}.   

 The transformation property of the vortices can be determined  from the moduli matrix \cite{Eto:2006pg}.
Indeed, the system possesses BPS saturated vortices described by the linearized equations
\be
\left(\D_1+i\D_2\right) \, q = 0,
\ee
\be
F_{12}^{(0)} + \frac{e^2}{2} \left( c \,{\bf 1}_N - q\, q^\dagger \right) =0; \qquad F_{12}^{(a)} + \frac{g_{N}^2}{2}\, q_{i}^\dagger  \, t^{a}\, q_{i}  =0.
\ee
The matter equation can be solved
\cite{Etou,Eto:2006pg}  ($z = x^1+ix^2$) by setting
\be
q  = S^{-1}(z,\bar z) \, H_0(z),\quad
A_1 + i\,A_2 = - 2\,i\,S^{-1}(z,\bar z) \, \bar\partial_z S(z,\bar z),
\ee
where $S$ is an  $N \times N$ invertible matrix, and  $H_{0}$ is  the  moduli matrix, holomorphic in $z$.  $S$ satisfies a simple second-order differential equation, which can be solved numerically. 

The individual vortex solution  breaks the color-flavor symmetry as
\be   SU(N)_{C+F} \to  SU(N-1) \times U(1),
\ee
leading  to the moduli space of the minimum vortices  which is
\be 
 {\cal M} \simeq  {\bf C}P^{N-1} = \frac{SU(N) }{ SU(N-1) \times U(1)}.
\ee
The vortex represented by the moduli matrix    (we consider here the vortices of minimal winding,  $k=1$)
\be   H_{0}(z)   \simeq  \left(\begin{array}{cccc}   1 & 0 & 0 & -a_1 \\   0 & \ddots & 0 & \vdots \\  0 & 0 & 1 & - a_{N-1} \\   0 & \ldots  & 0 & z\end{array}\right),
\label{minSUN}\ee
 can be shown explicitly \cite{Duality} to transform according to the fundamental representation of $SU(N)$.

\section{Dynamical Abelianization} 

 The fluctuations of internal, $CP^{N-1}$ moduli  in the above system are described in terms of a  $(2,2)$ supersymmetric
two-dimensional $CP^{N-1}$ sigma model \cite{ABEKY,Tong,SY}, and the known results in this model assure that in the long-distance limit  the system loses its orientation, and in the infrared there appear $N$  vortex ``vacua''  (analogue of the Witten's index).    Kinks appear connecting the different (vortex) vacua, which can be interpreted as the Abelian monopoles whose spectrum match nicely that found in the $4D$  theory \cite{HT,Tong,SY,GSY}.   
It means that the ``non-Abelian'' vortex found in the $U(N)$ model dynamically Abelianize, 
and this is not the type of system we are seeking.

As the concept of dynamical Abelianization is quite central to our discussion, and as this point is 
somewhat subtle, let us make a pause of reflection.  
  Dynamical Abelianzation, as normally understood, concerns the {\it gauge} symmetry. 
It means by definition that a non-Abelian gauge symmetry of a given theory  reduces at low energies by quantum effects to an Abelian ({\it dual or not})  gauge theory.  Related concepts are dynamical Higgs mechanism, or tumbling \cite{Tumbling}.   
Example of the theories in which this is known to occur are the pure ${\cal N}= 2$ supersymmetric Yang-Mills theories \cite{SW1,curves}   which reduce to Abelian gauge theories at low energies, and  the  $SU(2)$
 ${\cal N}= 2$  theories with $N_{f}=1,2,3$ matter hypermultiplets \cite{SW1}. 
 But as has been emphasized repeatedly and in Introduction above, ${\cal N}= 2$  supersymmetric $SU(N)$ QCD (with $N\ge 3$) {\it  with quark multiplets},  do not Abelianize in general \cite{APS,CKM,Konishi04}.   Whether or not the standard QCD  with light quarks Abelianizes is not known.  The 't Hooft-Mandelstam scenario implies a sort of dynamical Abelianization,  as it assumes the Abelian $U(1)^{2}$ monopoles to  be the dominant degrees of freedom at some relevant scales, but this has not been proven. 
 
    As the vortex orientation fluctuation modes are intimately connected to the way {\it dual } gauge symmetry emerges at low-energies \cite{Duality,Konishi04},  it is perfectly reasonable to use the same terminology for the vortex modes.
 
    Nevertheless, one could  {\it define}   the concept of {\it non-Abelian or Abelian vortices},  independently of the usual meaning attributed to it in relation to a gauge symmetry.  A vortex is non-Abelian, if it carries a non-trivial, internal  non-Abelian moduli, which can fluctuate along its length and in time.  We exclude from this consideration other vortex moduli   associated with their (transverse) positions, shapes  or sizes (in the case of higher-winding \cite{Hashimoto,AUSY,seven}  or semi-local vortices \cite{SYSemi,SemiSeven}).     Otherwise, a vortex is Abelian.  The standard ANO vortex is Abelian, as it possesses  no-continuous moduli.  The vortices found in the context of $U(N)$  models \cite{HT,ABEKY}   {\it are } indeed non-Abelian in this sense. 
    
    But just as a non-Abelian gauge group may or may not Abelianize depending on dynamics, a non-Abelian vortex may or may not dynamically Abelianize.  In the very papers in which these vortices have been discovered \cite{ABEKY,HT}  and in those which followed \cite{SY},  it was shown that they dynamically reduced to Abelian, ANO like vortices at long distances. The orientational moduli  fluctuate strongly and at long distances they effectively lose their orientation.  A recent observation \cite{SYLues}  nicely  exhibits this aspect through the L\"uscher term of the string tension.  
    
%
        
In a very recent paper \cite{DKO} it was shown that this fate is not unavoidable.   Semi-classical 
non-Abelian vortices which remain so at low-energies do exist;  they can be found in appropariate vacua, selected by a careful tuning of the bare quark masses. This is quite similar to the situation in ${\cal N}=1$ supersymmetric QCD, where a vacuum with a prescribed chiral  symmetry breaking pattern can be selected out of the degenerate set of vacua by appropriately tuning the bare quark mass ratios, before sending them to zero. The symmetry breaking pattern in those theories is aligned with the bare quark masses, as is well-known \cite{Kanomaly}.

The construction of vortices which do not completely Abelianize closes the gap in matching the results in the $4D$ theories at fully quantum regimes (where all bare mass parameters are small) and those in semi-classical regimes where the vortices can be reliably studied.  In other words the result of this work allows us to identify the semi-classical origin of the quantum non-Abelian monopoles found in \cite{APS,CKM}.   

\section{Non-Abelian vortices which do not dynamically reduce to ANO vortices } 

The model on which we shall base our consideration  is the softly broken ${\cal N}=2$  supersymmetric QCD with $SU(N)$ and $N_{f}=N$ flavors of quark multiplets,  
\bqa
&&  {\cal L}=     \frac{1}{ 8 \pi} {\im} \, {\tau}_{cl} \left[\int d^4 \theta \,
\Tr\,(\Phi^{\dagger} e^V \Phi e^{-V}) 
+\int d^2 \theta\,\frac {1}{ 2}\Tr\,(W W)\right]  \non \\
&& + {\cal L}^{({\rm quarks})}  +  \int \, d^2 \theta \,\mu  \,\Tr \,  \Phi^2;
\non 
\eea
where the quark Lagrangian is 
\bqa  &&  {\cal L}^{({\rm quarks})}  \non \\
&&   =  \sum_i \, \left[ \int d^4 \theta \,  ( Q_i^{\dagger} \, e^V \, 
Q_i + {\tilde Q_i} \,  e^{-V}\,  {\tilde Q}_i^{\dagger} ) +  \int d^2 \theta
\,  ( \sqrt{2}\,  {\tilde Q}_i \, \Phi \, Q^i    +      m_{i} \,    {\tilde Q}_i \, Q^i   )  \right]\;.
\non \eea
where  $\tau_{cl} \equiv  {\theta_0 / \pi} + {8 \pi i }/{ g_0^2}$ contains the coupling constant and the theta parameter,  $\mu$ is the adjoint scalar mass,  breaking softly  ${\cal N}=2$ supersymmetry to ${\cal N}=1$.    This is the same class of  theories as that of Section~\ref{aba:sec1},  but 
this time  we tune the bare quark masses as 
\[  m_{1}= \ldots = m_{n} = m^{(1)}; \quad m_{n+1}= m_{n+2}= \ldots = m_{n+r}= m^{(2)}\, ,
 \] 
\be  N=n+r\;;  \qquad    n\, m^{(1)} +  r\, m^{(2)} =0\;, \label{masses} \ee
or
\be   m^{(1)}=  \frac{r\, m_{0}}{\sqrt{ r^{2} + n^{2}}}, \quad   m^{(2)}= - \frac{ n \, m_{0}}{\sqrt{ r^{2} + n^{2} }},
\label{masscond}\ee
and their magnitude is taken as 
 \be   |m_{0}| \gg  |\mu|  \gg  \Lambda\,.\label{masssemi}
\ee
  The adjoint scalar VEV can be taken to be 
\be \brc \Phi \ckt   =  - \frac{1}{\sqrt{2}}  \left(\begin{array}{cc}m^{(1)} \, {\mathbbm 1}_{n \times n} & {\bf 0} \\ {\bf 0}  & m^{(2)} \, {\mathbbm 1}_{r \times r}  \end{array}\right)  \label{ourvac}
\ee
Below the mass scale  $v_{1} \sim   |m_{i}|$   the system thus  reduces to a gauge theory with gauge group 
\be   G =   \frac {SU(n) \times SU(r) \times U(1)}{{\mathbbm Z}_{K}}\,,   \quad K = {\rm LCM} \, \{n,r\}\;    \label{Gaugegr} 
\ee
 where $K$ is the least common multiple of $n$ and $r$.   The higher $n$ color components of the first $n$ flavors  (with the bare mass $m^{(1)}$)   remain massless, as well as the lower $r$ color components of the last $r$ flavors
 (with the bare mass $m^{(2)}$):  they will be denoted as $q^{(1)}$ and $q^{(2)},$
 respectively.   
Our model  then is: 
\bqa  && {\cal L} =  - \frac{ 1 }{4 g_{0}^2}  F_{\mu \nu}^{0\, 2}  - \frac{ 1 }{4 g_{n}^2}  F_{\mu \nu}^{n\, 2} - \frac{ 1 }{4 g_{r}^2}  F_{\mu \nu}^{r\, 2}   +    \frac { 1}{ g_{0}^2}  |{\cal D}_{\mu} \Phi^{(0)}|^2 +  \frac { 1}{ g_{n}^2}  |{\cal D}_{\mu} \Phi^{(n)}|^2 +  \non \\
&& +   \frac { 1}{ g_{r}^2}  |{\cal D}_{\mu} \Phi^{(r)}|^2  +   \left|{\cal D}_{\mu}  q^{(1)} \right|^2 + \left|{\cal D}_{\mu} \bar{\tilde{q}}^{(1)}\right|^2  + 
  \left|{\cal D}_{\mu}  q^{(2)} \right|^2 + \left|{\cal D}_{\mu} \bar{\tilde{q}}^{(2)}\right|^2     -    V_D -  V_F, 
\non \eea
plus fermionic terms,  
where  $V_{D}$ and $V_{F}$ are the $D$-term and $F$-term potentials.  
%
The light squark fields are the components
\be   Q(x) =   \left(\begin{array}{c c}  q^{(1)}(x) _{n\times n}  & 0 \\ 0  &  q^{(2)}(x)_{r \times r}\end{array}\right)\,,\quad   {\tilde Q}(x) =   \left(\begin{array}{c c}  {\tilde q}^{(1)}(x) _{n\times n}  & 0 \\ 0  &  {\tilde q}^{(2)}(x)_{r \times r}\end{array}\right)\,, \ee
if written in a color-flavor mixed matrix notation.   
We keep 
\be   {\tilde{q}}^{(1)}  =  (q^{(1)})^{\dagger}, \qquad  {q}^{(2)}  =  -  ({\tilde q}^{(2)})^{\dagger}\;;
\label{reduction1}   \ee
the redefinition   
\be  q^{(1)} \to \frac{1}{\sqrt{2}} \, q^{(1)}, \quad  {\tilde q}^{(2)} \to \frac{1}{\sqrt{2}} \, {\tilde q}^{(2)}
\label{reduction2}   \ee
 brings the kinetic terms for these fields back to the original  form. 

The VEVs of the adjoint scalars are  given by 
$ \brc \Phi^{(0)} \ckt =-  m_{0}, \quad \brc \Phi^{(a)} \ckt = \brc \Phi^{(b)} \ckt   =0,$   while the squark VEVs are given  by
\be    
\brc Q \ckt =   \left(\begin{array}{c c}  v^{(1)} \, {\mathbbm 1}_{n\times n}  & 0 \\ 0  &  - v^{(2)\, *} \, {\mathbbm 1}_{r\times r}  \end{array}\right)\,,\quad \brc {\tilde Q} \ckt =   \left(\begin{array}{c c}  v^{(1)\, *} \, {\mathbbm 1}_{n\times n}  & 0 \\ 0  &     v^{(2)} \, {\mathbbm 1}_{r\times r}  \end{array}\right)\,,
\label{QkVEV}\ee
with
\be 
 | v^{(1)}|^{2} +    | v^{(2)}|^{2}  = \sqrt{{n+r}/ {n\, r}} \,  \mu \, m_{0}\,  \;. 
\ee
There is a continuous vacuum degeneracy;  we take 
\[    v^{(1)} \ne 0; \qquad  v^{(2)} \ne 0\;,
\]
in the following.  

\begin{figure}
\begin{center}
\includegraphics[width=2.5 in]{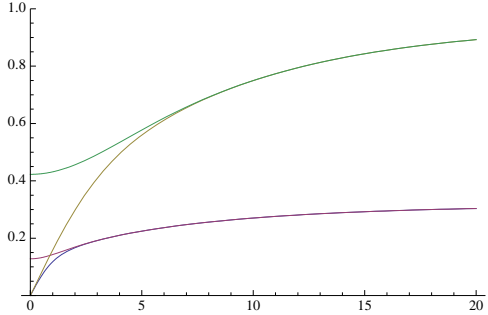}
\caption{\footnotesize Numerical result for the profile functions $f_{1,2}, g_{1,2}$ as functions of the radius $\rho$, for $SU(3)\times SU(2)\times U(1)$  theory.  The coupling constants and the ratio of the VEVs are taken to be $g_{0}=0.1$, $g_{3}=10$, $g_{2}=1$, $|v_{2}|/ |v_{1}|=3$.  }
\label{Vortexprofile}
\end{center}
\end{figure}

The vacuum breaks the gauge group $G$ completely, leaving at the same time a color-flavor diagonal symmetry 
\be   [ SU(n) \times SU(r) \times U(1) ]_{C+F}  
\label{colorflavor}\ee
 unbroken.   The full global symmetry, including the overall global $U(1)$ is given by 
$   U(n) \times U(r)\;.    $
   The minimal vortex in this system 

 has  e.g.   the form,  
\be   q^{(1)}  =   \left(\begin{array}{c c    }  e^{i \phi} \,f_{1}(\rho)  &  0   \\
0 &   f_{2}(\rho)\, {\mathbbm 1}_{(n-1)\times (n-1)}  
 \end{array}\right)\,, \quad 
 {\tilde q}^{(2)}  =   \left(\begin{array}{c c    }  e^{i \phi} \,g_{1}(\rho)  &  0   \\
0 &  g_{2}(\rho)\, {\mathbbm 1}_{(r-1)\times (r-1)}  
 \end{array}\right)\,, 
\label{vortexconf} \ee
 where $\rho$ and $\phi$ stand for the polar coordinates in the plane perpendicular to the vortex axis,  $f_{1,2}, g_{1,2}$ are profile functions.  
  The behavior of numerically integrated vortex profile functions $f_{1,2}, g_{1,2}$ is illustrated in Fig.~\ref{Vortexprofile}. 

   We note here only that the necessary boundary conditions on the squark profile functions have the form, 
   \[     f_{1}(\infty)=  f_{2}(\infty) = v^{(1)}, \qquad  g_{1}(\infty)= g_{2}(\infty) = v^{(2)}, 
   \]
   while at the vortex core, 
   \be   f_{1}(0)=0, \quad  g_{1}(0) =0, \qquad   f_{2}(0) \ne 0, \quad  g_{2}(0) \ne 0,
  \label{fundamental} \ee
  
  The most important fact about these minimum vortices  is that  one of the $q^{(1)}$ {\it and}  one of the ${\tilde q}^{(2)}$ fields must necessarily wind at infinity, simultaneously.  
   As the individual vortex breaks the (global) symmetry of the vacuum as 
\be  [ SU(n) \times SU(r) \times U(1) ]_{C+F}  \to  SU(n-1)  \times SU(r-1)  \times U(1)^{3}, 
\label{smaller}  \ee
the vortex acquires  Nambu-Goldstone modes parametrizing
\be   CP^{n-1} \times CP^{r-1}\;:         
\label{nambu}\ee
they transform under the exact color-flavor symmetry $SU(n) \times SU(r)$  as the 
bi-fundamental  representation, $({\underline n}, {\underline r})$.
Allowing the vortex orientation to fluctuate along the vortex length and in time, we get a $CP^{n-1} \times CP^{r-1}$  two-dimensional sigma model as an effective Lagrangian describing them.  The details have been worked out in \cite{SY,HT}  and  need not be repeated here.

The main  idea  is this.  Let us assume without losing generality that  $n > r$ (excluding the special  case of $r=n$).   As has been shown in \cite{SY,HT} the coupling constant of the $CP^{n-1}$  sigma models grows precisely as the  coupling constant of the $4D$ $SU(n)$ gauge theory.   At the point the $CP^{n-1}$  vortex moduli fluctuations become strong and 
the dynamical scale $\Lambda$ gets generated, with vortex kinks (Abelian monopoles) acquiring mass of the order of  $\Lambda$, 
the vortex still carries the unbroken $SU(r)$  fluctuation modes ($CP^{r-1}$),  as 
the $SU(r)$ interactions are still weak.  
  Such a vortex will carry one of the $U(1)$ flux 
arising from the dynamical breaking of $SU(n) \times U(1) \to  U(1)^{n}$, as well as an $SU(r)$ flux.  
As these vortices end at a massive monopole (arising from the high-energy gauge symmetry breaking, Eq.~(\ref{ourvac})),  the latter necessarily carries a non-Abelian continuous moduli, whose points transform as in
the fundamental representation of 
$SU(r)$.  This can be  interpreted as the  (electric description of)  dual gauge $SU(r)$ system observed in the infrared limit of the $4D$ SQCD \cite{APS,CKM}.  
\begin{figure}
\begin{center}
\includegraphics[width=3 in]{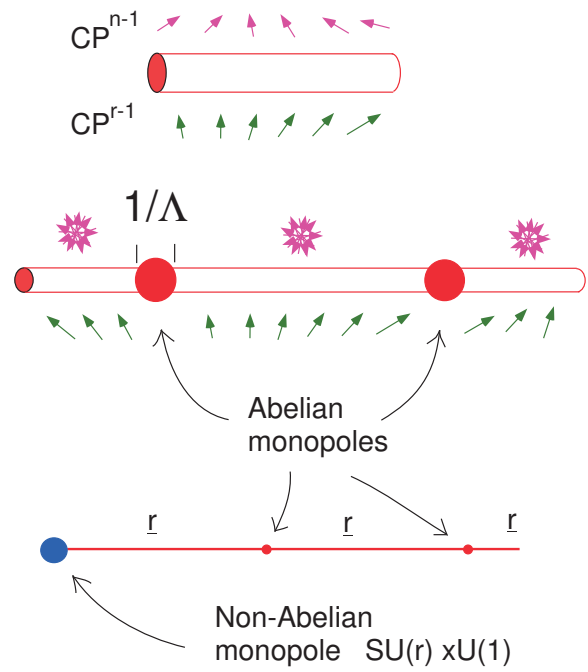}
\caption{\small The vortex with  $CP^{n-1}\times CP^{r-1}$ orientational modes.
}
\label{twovor}
\end{center}
\end{figure} 
  
     Thus vortices with non-Abelian moduli,  which do not dynamically Abelianize completely, can be constructed in a natural way. 
 Semi-classically, they are simply vortices carrying the $SU(n) \times SU(r) \times U(1)$ color-flavor flux.  More precisely,  they carry the Nambu-Goldstone modes 
Eq.~(\ref{nambu})   resulting from the partial breaking of the $SU(n) \times SU(r) \times U(1)$  global symmetry by the vortex.   For  $n>r$,   $CP^{n-1}$ field fluctuations propagating along the vortex length  become strongly coupled in the infrared, the $SU(n) \times U(1)$ part dynamically Abelianizes;  the vortex however still carries weakly-fluctuating $SU(r)$ flux modulations.  In our  theory  where  $SU(n) \times SU(r) \times U(1)$ model emerges as the low-energy approximation of an underlying $SU(N)$ theory, such a vortex is not stable. When the vortex ends at a monopole, its  $CP^{r-1}$ orientational modes are turned into the dual $SU(r)$ color modulations of the monopole. 

\bibliographystyle{ws-procs9x6}

\bibliography{ws-pro-sample}

\end{document}